\documentclass{ws-mpla}
\begin{document}

\def\be{\begin{equation}}
\def\ee{\end{equation}}
\def\bea{\begin{eqnarray}}
\def\eea{\end{eqnarray}}
\def\E{{\rm e}}
\def\bearst{\begin{eqnarray*}}
\def\eearst{\end{eqnarray*}}
\def\peleven{\parbox{11cm}}
\def\peffec{\peight{\bearst\eearst}\hfill\peleven}
\def\pspace{\peight{\bearst\eearst}\hfill}
\def\ptwelve{\parbox{12cm}}
\def\peight{\parbox{8mm}}
\def\al{&\!\!\!\!}
\def\x{{\bf x}}
\def\f{\frac}
\def\a{\alpha}

\markboth{Rahvar \& Sobouti} {An Inverse $f(R)$ Gravitation for
Cosmic Speed up, and Dark Energy Equivalent}

\catchline{}{}{}{}{}

\title{An Inverse $f(R)$ Gravitation for Cosmic Speed up, and
 Dark Energy Equivalent }

\author{Sohrab Rahvar}
\address{Department of Physics, Sharif University of
Technology, P.O.Box 11365-9161, Tehran,
Iran.\footnote{rahvar@sharif.edu}}

\author{Yousef Sobouti}

\address{Institute for Advanced Studies in Basic
Sciences,\\
 P.O.Box 45195-1159, Zanjan, Iran}

\maketitle

\pub{Received (Day Month Year)}{Revised (Day Month Year)}

\begin{abstract}
To explain the cosmic speed up, brought to light by the recent SNIa
and CMB observations, we propose the following: a) In a spacetime
endowed with a FRW metric, we choose an empirical scale factor that
best explains the observations. b) We assume a modified gravity,
generated by an unspecified field lagrangian, $f(R)$. c) We use the
adopted empirical scale factor to work back retroactively to obtain
$f(R)$, hence the term `Inverse $f(R)$'. d) Next we consider the
classic GR and a conventional FRW universe that, in addition to its
known baryonic content, possesses a hypothetical `Dark Energy'
component.  We compare the two scenarios, and find the density, the
pressure, and the equation of the state of the Dark Energy required
to make up for the differences between the conventional and the
modified GR models.
\keywords{Cosmology; Dark Energy; Modified
Gravity.}

\end{abstract}
\ccode{95.36.+x, 98.80.Jk, 98.80.Es}

\hspace{.3in}
\newpage

As cosmological standard candles, supernovae type Ia (SNIa) appear
dimmer than what one expects from a Cold Dark Matter (CDM) model of
the universe \cite{acc1,acc2,acc3}. This observation and other
evidences from the Cosmic Microwave Background (CMB) measurements
indicate that the universe is in an acceleration phase of its
expansion \cite{Ben1,Ben2,Ben3}. A conventional CDM scenario does
not explain this speed up. Some authors have stipulated a dark
energy component to make up for whatever dynamical effects that the
known energy momentum content of the model leaves unaccounted for
\cite{de1,de2,de3,de4,de5,de6,de7,de8,de9,de10,de11,de12,de13,de14,de15}.
Others have entertained alternatives to Einstein's gravitation
\cite{mg1,mg2,mg3,mg4,mg5,mg6,mg7,mg8,mg9}. Yet a third school have
resorted to inhomogeneous FRW universes to explain the dilemma
\cite{cel07}. Since all these approaches attempt to answer the same
question, all should be equivalent, and there should be a way to
translate one language to the other.

Here, we are concerned with the 'dark energy' and 'alternative
gravitation' scenarios. We suggest to begin with a Freidman-
Robertson- Walker (FRW) universe, to choose its scale factor,
$a(t)$, in a way that best explains the available observations,
and to work out the dynamics of the spacetime. Next, to write down
the field equations for a modified $f(R)$ gravitation, and knowing
$a(t)$, to solve for $f(R)$. Finally, to attribute whatever
deviations from the conventional FRW results there is, to a dark
energy field, and to obtain its density, pressure, equation of state, etc.\\

\begin{figure}[h]
\centerline{\psfig{file=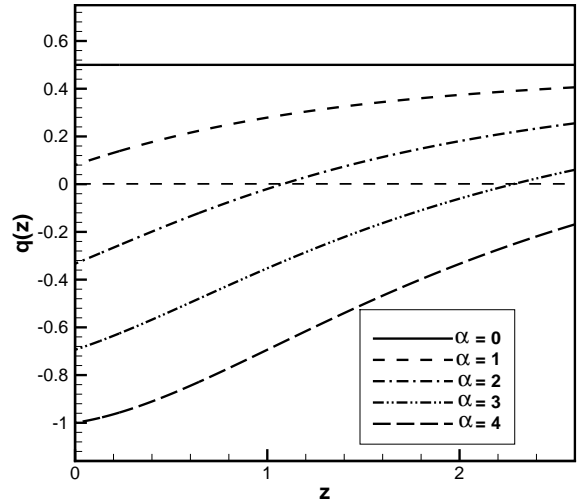,width=3.5in}} \vspace*{8pt}
\caption{Plot of $q(t)$ versus the redshift, $z=a(t_0)/a(t)-1$, for
$\alpha=0, 1, 2, 3, 4$; and $p=1/3$. The case $\alpha=0$ gives the
classic value, $q=2/3$. As $\alpha$ increases from 1 to 4,
transition to the accelerated phase of expansion moves from later to
earlier epochs, from smaller $z$'s to larger one.}\label{f1}
\end{figure}

\noindent\textbf{The choice of the scale factor}\\
\noindent In a FRW metric, $ds^2=-dt^2+a(t)^2 d\textbf{x}^2$, single
term scale factors of the form $a\propto t^{\beta}$ lead to constant
deceleration parameters, $q=-\ddot{a}a/\dot{a}^2=(1-\beta)/\beta$,
and do not serve the purpose. We propose the following two-term
ansatz
\begin{eqnarray}\label{e1}
\al\al a(t)=\frac{1}{1+p}(t/t_0)^{2/3}\left[1+p
(t/t_0)^{2\alpha/3}\right],
\end{eqnarray}\vspace{1mm}
\noindent where $t_0$ is the age of the universe, and $\alpha$ and
$p$ are the free parameters of the model, to be adjusted to ensure
compatibility of the emerging results with observations. The
factor $(1+p)^{-1}$ is introduced to have $a(t_{0})=1$. By letting
either $\alpha$ or $p$ tend to zero, one recovers the standard CDM
universe. Hereafter, for economy in writing, we will use the time
parameter $\tau= p(t/t_0)^{2\alpha/3}$ instead of the conventional
time $t$. From Eq. (\ref{e1}) one finds
\begin{eqnarray}\label{e2}
q\al~=~\al \f{1}{2}\left[1-(1+\alpha)(2\alpha-1)\tau\right]
\left[1+\tau\right]\left[1+(1+\alpha)\tau\right]^{-2}\\
\label{e3} H\al~=~\al\f{2}{3t_0}(\f{\tau}{p})^{-3/2\alpha}
\left[1+(1+\alpha)\tau\right]\left[1+\tau\right]^{-1},\\
\label{e4} {\cal R}\al~=~\al{t_0}^2R=
6H^2(1-q)=\f{4}{3}(\f{\tau}{p})^{-3/\alpha}\\\nonumber ~\al
\times~\al [1+(2+5\alpha+2\alpha^2)\tau
+(1+5\alpha+4\alpha^2)\tau^2] \left[1+\tau\right]^{-2}.
\end{eqnarray}
For $\a>1/2$, $q$ can become negative and remain nonsingular
throughout. Transition from a decelerated phase of expansion to an
accelerated one takes place at
$\tau_{\rm{trans}}=\left[(\alpha+1)(2\alpha-1)\right]^{-1}$ or
$t_{\rm{trans}}=\left[(\alpha+1)(2\alpha-1)p\right]^{-3/2\alpha}t_0<t_0$.
For $-1<\a<1/2$, $q$ is always positive and nonsingular. For
$\a<-1$, q can become negative but also singular. The last two
possibilities are discarded. For all values of $\a$ and $p$,
\begin{eqnarray} \label{e6}
\al\al\rm{limit} ~q(t\rightarrow 0)=\f{1}{2},~ \nonumber\\
\al\al\rm{limit} ~ q(t\rightarrow \infty)=(1-2\a)/2(1+\a).
\end{eqnarray}
In Fig.(\ref{f1}) we have plotted $q(t)$ versus the redshift,
$z=a(t_0)/a(t)-1$, for several values of $\alpha$ and $p=1/3$.
\\
As to $H$  and $R$, both remain positive for all times. Both tend
to $\infty$ as $\tau \rightarrow 0$ and decease monotonically to
$0$ as $\tau \rightarrow \infty$. They exhibit a normal behavior
in the neighborhood of
$\tau_{\rm{trans}}=[(1+\alpha)(2\alpha-1)]^{-1}$.

 Equation (\ref{e3}), written for the present epoch, reveals a
 constraint on $\alpha$ and $p$, that should be observed in the final
 design of the model. Thus,

\begin{eqnarray} \label{e7}
[1+(1+\alpha)p]/[1+p]= \f{3}{2}H_0 t_0\approx\f{3}{2}.
\end{eqnarray}\\

\begin{figure}
\centerline{\psfig{file=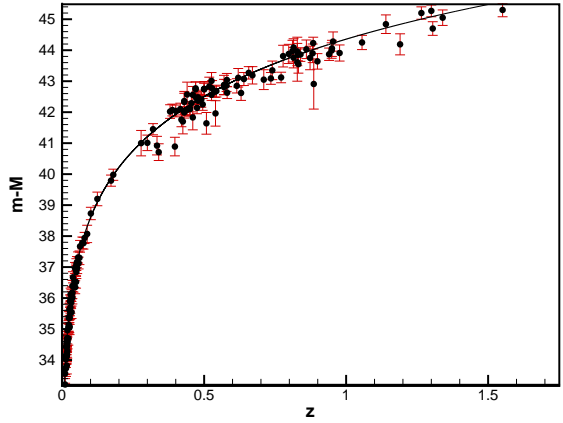,width=3.5in}} \vspace*{8pt}
\caption{The distance modulus of the SNIa Gold's sample versus
redshifts, black circles; And the plot of Eq. (\ref{e8}) with the
scale factor of Eq. (\ref{e1}), solid line. Parameters of the model
are $\alpha = 2$ and $p=1/3$.}\label{f2}
\end{figure}

\noindent\textbf{Empirical values of $\a$ \& $p$ }\\
The distance modulus (corrected for the reddening)  and the
(dimensionless) luminosity distance, $D_L(z;\a,p)$, of supernovae
are related as
\begin{eqnarray} \label{e8}
\mu= m-M~~\al=\al~~ 5\log D_L (z; \alpha,p)+25, \\\nonumber
D_L(z;\alpha, p) \al~~=~~\al (1+z)\int dz{H(z;\alpha, p)^{-1}}.
\end{eqnarray}
In Fig. (\ref{f2}), the observed distance modulus of 157 SNIa of
Gold's sample are plotted versus the redshifts. The solid curve is
the plot of equation (\ref{e8}), in which, in compliance with the
constraint of Eq. (\ref{e7}), we have chosen
\begin{eqnarray} \label{e9}
\al\al \a=2,~~~~~p=1/3.
\end{eqnarray}
The fit to the data points is adequate for our purpose, though the
parameters can be refined to optimize the fit. These numbers give
$t_{\rm{trans}} = 0.43 t_0$ and $z_{\rm{trans}} = 1.14$. On
adopting $H_0\approx 70 \rm{km ~sec^{-1}~Mpc^{-1}}$, one obtains
an age of $t_0 \approx 12.4 \rm{Gyr}$ for the universe.
\\

\begin{figure}
\centerline{\psfig{file=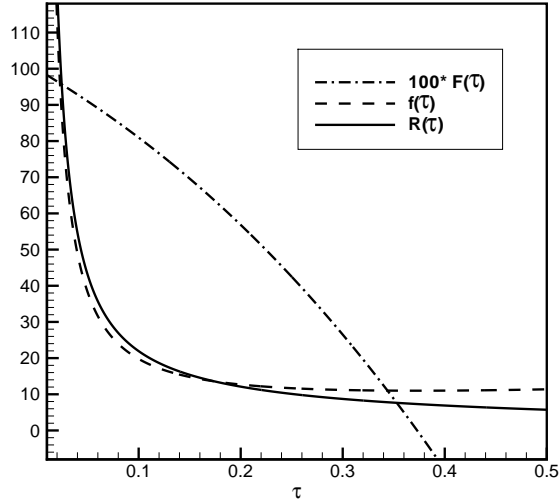,width=3.5in}} \vspace*{8pt}
\caption{$F(\tau)$, proper solution of Eq. (\ref{e14})
($\times100$), dot- dashed line; $f(\tau)$, dashed line; and ${\cal
R}(\tau)$, solid line; $\alpha=2$, $p=1/3$.}\label{f3}
\end{figure}

\noindent\textbf{An inverse $f(R)$ way out}\\
\noindent We begin with a modified field equation generated by an,
as yet, unspecified field lagrangian, $f(R)$,
\begin{eqnarray} \label{e10}
R_{\mu\nu} \al~-~\al \frac{1}{2}Rg_{\mu\nu}=\frac{1}{2F}(f-RF)g_{\mu\nu}\\
\al~+~
\al\frac{1}{F}(\nabla_{\mu}\nabla_{\nu}-g_{\mu\nu}\nabla_{\lambda}\nabla^{\lambda})F-
\frac{\kappa}{F} T^{(M)}_{\mu\nu},\nonumber
\end{eqnarray}
where $F(R)=d f(R)/dR$, and $\kappa=8\pi G$. For a universe of FRW
type, filled with a perfect fluid of density $\rho_m$ and
pressure $p_m$, Eq. (\ref{e10}) and the equation of continuity
reduce to
\begin{eqnarray}\label{e11}
\al\al 3H\dot F+ 3H^{2}F-\frac{1}{2}(RF-f)-\kappa\rho_{m}=0,\\
\al\al\label{e12} \ddot{F}-H\dot{F}+2\dot{H}F+\kappa(\rho_m+p_m)=0,\\
\al\al\label{e13}\dot{\rho}_m+3H(\rho_m+p_m)=0,
\end{eqnarray}
We further neglect the pressure and integrate Eq. (\ref{e13}) to
obtain $\rho_m(t)=\rho_0 a(t)^{-3}$. Next we substitute for
$\rho_m$ in Eq. (\ref{e12}), assume $\alpha=2$, change the time
variable to $\tau=p(t/t_0)^{4/3}$, and find
\begin{eqnarray}\label{e14}
\al\al(1+\tau)^3 \tau^2 F^{\prime\prime}- \f{1}{4}(1+\tau)^2(1+5\tau) \tau F^\prime\\
\al\al~~~-\f{4}{3}(1+\tau)(1+\f{3}{4}\tau+3\tau^2)F+\f{4}{3}=0,\nonumber
\end{eqnarray}
where the $'\prime~'$ now stands for $d/d\tau$, and we have,
arbitrarily, put the dimensionless constant $(3/4)^3(1+p)^3{t_0}^2
\kappa \rho_0$, that appear in the course of mathematical
manipulations equal to $one$. We will shortly discuss the
numerical solution of Eq. (\ref{e14}). Some general remarks on its
asymptotic behavior, however, are instructive. As
$\tau\rightarrow0$ we find
\begin{eqnarray}\label{e15}
\al\al F(\tau)=
\left[1-\f{7}{4}\tau-\f{23}{6}\tau^2+\dots\right]\\\nonumber
\al\al~~~+c_1~\tau^{-0.44}P_1(\tau)+c_2~\tau^{1.7}P_2(\tau),~\tau\rightarrow0.
\end{eqnarray}
where $P_1 $ and $ P_2$ are calculable polynomials in $\tau$, and
begin with term 1; $c_1$ and $c_2$ are constants of integration to
be obtained from boundary conditions. The exponents, $-0.44 $ and
$1.7$ are approximate solutions of the indicial equation,
$s^2-\f{5}{4}s -\f{3}{4} =0$. As $\tau\rightarrow0$, $F$ diverges to
$\infty$ or converge to $0$ on account of one or the other term.
This feature makes the solutions sensitive to a CDM type boundary
conditions of the form, $F(\tau_{\rm{initial}})=1$ and
$F^\prime(\tau_{\rm{initial}})=0$. Presently we have no basis,
observational or otherwise, to make an intelligent guess as to what
the appropriate boundary conditions are. For the sake of argument,
however, we have adopt $c_1=c_2=0$,  and kept only the proper
solution of Eq. (\ref{e13}). With $F(\tau)$ known, Eq.(\ref{e11})
becomes an algebraic equation to calculate $f(\tau)$. The
numerically calculated solutions of $F(\tau)$, $f(\tau)$ and ${\cal
R}={t_0}^2 R(\tau)$ are plotted in Fig. (\ref{f3}). Elimination of
$\tau$ in favor $R$ provides $F(R)$ and $f(R)$

It is instructive to examine the asymptotic behavior of  $F(R)$
and $f(R)$ analytically.  In the limit of small and large
$\tau$'s, corresponding to large and small $R$'s, one finds
\begin{eqnarray} \label{e16}
F(R) \al~=~\al \left[1-\f{7}{6}\sqrt[3]{6}p{\cal
R}^{-2/3}+\cdots\right]~~~{\cal R}\rightarrow\infty\\
\label{e17}\al~=~\al-\frac{11}{4}\left(\f{{\cal
R}}{p}\right)\left[1+\frac{5}{28\sqrt[3]{3}}\left(\f{{\cal
R}}{p}\right)^{2/3}\right]~~{\cal R}\rightarrow0.
\end{eqnarray}
Note that $F(R)$ is a dimensionless scalar as it should be. With
$F(R)=df/dR$ known, it is a matter of simple integration to
obtain $f(R)$. Thus,
\begin{eqnarray} \label{e18}
f(R) \al~=~\al \f{{\cal
R}}{{t_0}^2}\left[1+\frac{7}{4}\sqrt[3]{6}p{\cal
R}^{-2/3}+\cdots\right],~~~{\cal R}\rightarrow\infty,\\
\label{e19} \al ~=~\al -\frac{11}{8 {t_0}^2}\left(\f{{\cal
R}}{p}\right)^{2}\left[1+\frac{15}{112\sqrt[3]{3}}\left(\f{{\cal
R}}{p}\right)^{2/3}\right],~~~{\cal R}\rightarrow0.
\end{eqnarray}
\begin{figure}[h]
\centerline{\psfig{file=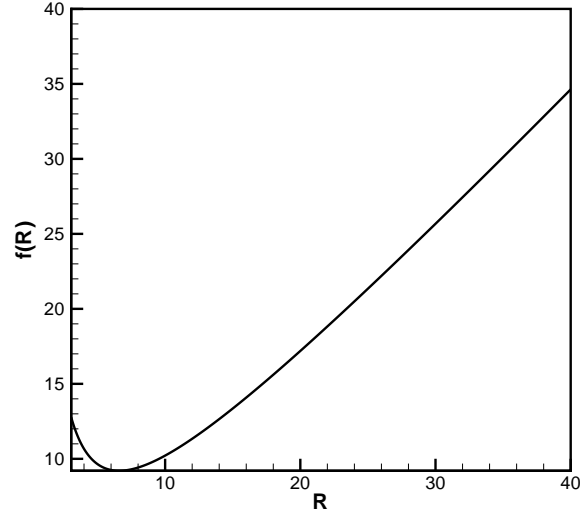,width=3.5in}} \vspace*{8pt}
\caption{Numerical plot of $f(R)$ versus $R$. $\tau$ is  eliminated
between $f(\tau)$ and $R(\tau)$; $\alpha=2$, $p=1/3$.}\label{f4}
\end{figure}
At early epochs (large $R$'s), the spacetime approaches the
conventional FRW universe with the classical GR while for the later
times we have a positive acceleration universe. Another point about
this action is that for $R\rightarrow 0$ in the solar system scales,
$f(0)= 0$ and $f'(0)=0$. In this case we will have standard GR
equation and $f(R)$ evades from the solar system test. Recently this
type of models have been studied by introducing action in the form
of $ f(R)= R + f_1(R)$, where for $R=0$ in the solar system,
$f(0)=0$ and for larger $R$'s, in cosmological scales and inside the
large scale
structures the action reduces to $f(R) = R-\Lambda$ \cite{star07,star07-2}. \\
In the remaining range of ${\cal R}$, integration is done
numerically and the results are plotted in Fig. (\ref{f4}).\\

\begin{figure}[h]
\centerline{\psfig{file=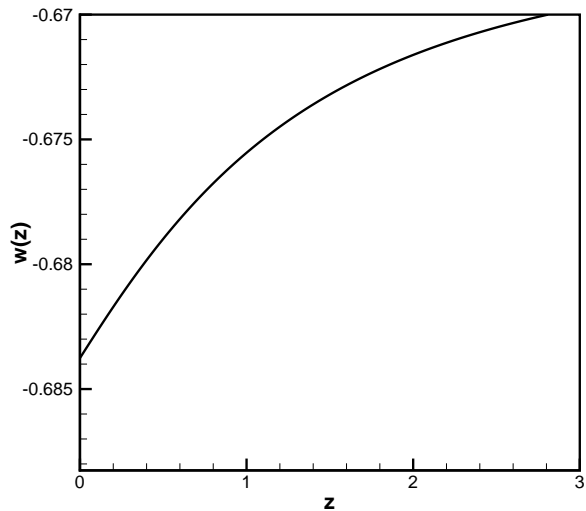,width=3.5in}} \vspace*{8pt}
\caption{Equation of State of the Dark Energy: Numerical plot of
 $w=p_{de}/\rho_{de}$ versus the redshift
}\label{f6}
\end{figure}

\noindent \textbf{Dark Energy equivalent}\\
Instead of the modified gravitation considered above, let us
assume a classic FRW universe. That is, let $f(R)=R$ and $F=1$.
Let this universe, however, have a 'Dark Energy' component, in
addition to its conventional baryonic content. The counterparts of
Eqs. (\ref{e11}), and (\ref{e12}) will be
\begin{eqnarray}
\label{e21}\al\al 3H^2-\kappa\left(\rho_{de}+\rho_{m}\right)=0,\\
\label{e22}\al\al 2\dot
H+\kappa\left(\rho_{de}+\rho_{m}\right)+\kappa\left(p_{de}+p_{m}\right)=0.
\end{eqnarray}
Subtracting Eq. (\ref{e11}) from (\ref{e21}), and Eq. (\ref{e12})
from (\ref{e22}) gives
\begin{eqnarray} \label{e23}
\al\al \kappa\rho_{de}=3H^2(1-F)-3H\dot F+\frac{1}{2}(R F-f),\\
\al\al \kappa(\rho_{de}+p_{de})=\ddot F-H \dot F-2\dot
H(1-F),~~\textrm{or}\nonumber\\
\label{e24} \al\al \cr \al\al \kappa p_{de}=\ddot F
+2H\dot F-H^2(1-2q)(1-F)\nonumber\\
\al\al~~~~~~~-\frac{1}{2}(R F-f). \label{e24}
\end{eqnarray}
The equation of state for the dark energy is obtained by
eliminating $\tau$, implicit in $F$ and $H$, between Eqs.
(\ref{e23}) and (\ref{e24}). This is done numerically and
$w=p_{de}/\rho_{de}$ as a function of the redshift is plotted
in Fig. (\ref{f6}).\\

\begin{figure}
\centerline{\psfig{file=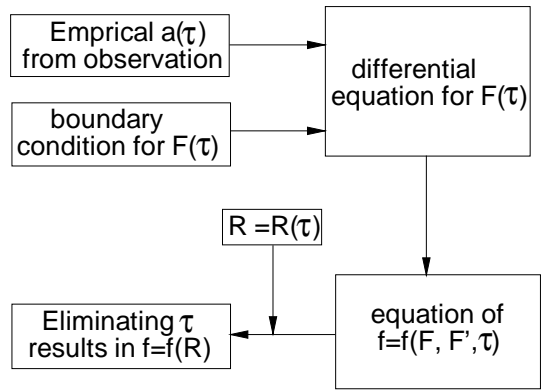,width=3.5in}} \vspace*{8pt}
\caption{Algorithm of inverse $f(R)$: Choose a scale factor; Solve
field equations first for $F=df/dR$ and next for $f$ as functions of
the time parameter $\tau$; Eliminate $\tau$ between $R(\tau)$
$f(\tau)$ to arrive at $f(R)$.}\label{alg}
\end{figure}

\noindent\textbf{Concluding remarks}\\
The algorithm of Fig. \ref{alg} summarizes the path we have
followed in this letter.
We have resorted to the SNIa observations to design an empirical
FRW metric that allows the model universe to transit from  a phase
of decelerated expansion at early epochs to an accelerated one at
later times. The spacetime is almost a CDM model and the gravity
is almost the classic GR at very early times, but evolves away in
course of time. Next we have maintained that  the so-designed
spacetime is deducible from a modified non- Hilbert- Einstein
field lagrangian, $f(R)$ . Knowing the metric, we have solved the
modified field equations retroactively for the sought-after
$f(R)$. Finally, we have compared our results with those of a
conventional FRW model and have attributed the differences between
the two to a dark energy component. Eventually, we have extracted
the density, the pressure, and the equation of
state of this stipulated energy.\\
We note that our choice of the scale factor and the adjustment of
its free parameters, to comply with the available cosmological
observations, is, by no means, unique. The goal is simply to
demonstrate that the use of the observations at the outset, to
deduce the rudiments of what seems reasonable, facilitates the
access to possible formal underlying theories, the action based
$f(R)$ gravity in our case. With the availability of more
extensive and more accurate data in future one may come back and
revise the model. See also \cite{arman} for a similar
emphasis.\\
One of us (YS)\cite{sob07} has followed the same path to propose a
modified gravitation for galactic environments and to explain the
flat rotation curves and the Tully-Fisher relation in spiral
galaxies without recourse to hypothetical dark matters.


\newpage
\end{document}